\documentclass[twocolumn,showpacs,showkeys,preprintnumbers,superscriptaddress,amsmath,amssymb]{revtex4-1}
\usepackage[english]{babel}
\usepackage{amsmath}
\usepackage{graphicx}
\usepackage{float}

\usepackage{siunitx}

\usepackage[normalem]{ulem}
\usepackage[dvipsnames,usenames]{xcolor}

\usepackage[version=3]{mhchem} 

\usepackage{dcolumn}
\usepackage{times}

\newcommand{\bcs}{\ce{BaCoS2}}
\newcommand{\bcns}{\ce{BaCo$_{1-x}$Ni$_x$S2}}

\newcommand{\sigw}{\ensuremath{\sigma(\omega)}}
\newcommand{\resigw}{\ensuremath{\sigma_1(\omega)}}

\newcommand{\K}[1]{\SI{#1}{\kelvin}}
\newcommand{\mev}[1]{\SI{#1}{\milli\electronvolt}}
\newcommand{\ev}[1]{\SI{#1}{\electronvolt}}

\begin{document}

\title{Linear behavior of the optical conductivity and incoherent charge transport in \bcs}

\author{D. Santos-Cottin}
\affiliation{LPEM, ESPCI Paris, PSL University; CNRS; 10 rue Vauquelin, F-75005 Paris, France}
\affiliation{Sorbonne Universit\'e; CNRS; LPEM;  F-75005 Paris, France}

\author{Y. Klein}
\affiliation{Institut de Min\'eralogie, de Physique des Mat\'eriaux et
  de Cosmochimie (IMPMC), Sorbonne Universit\'e, CNRS UMR 7590, IRD
  UMR 206, MNHN; F-75252 Paris, France}

\author{Ph. Werner}
\affiliation{Department of Physics, University of Fribourg, 1700 Fribourg, Switzerland}

\author{T. Miyake}
\affiliation{Research Center for Computational Design of Advanced Functional Materials, AIST, Tsukuba 305-8568, Japan}

\author{L. de' Medici}
\affiliation{LPEM, ESPCI Paris, PSL University; CNRS; 10 rue Vauquelin, F-75005 Paris, France}
\affiliation{Sorbonne Universit\'e; CNRS; LPEM;  F-75005 Paris, France}

\author{A. Gauzzi}
\affiliation{Institut de Min\'eralogie, de Physique des Mat\'eriaux et
  de Cosmochimie (IMPMC), Sorbonne Universit\'e, CNRS UMR 7590, IRD
  UMR 206, MNHN; F-75252 Paris, France}

\author{R. P. S. M. Lobo}
\affiliation{LPEM, ESPCI Paris, PSL University; CNRS; 10 rue Vauquelin, F-75005 Paris, France}
\affiliation{Sorbonne Universit\'e; CNRS; LPEM;  F-75005 Paris, France}

\author{M. Casula}
\affiliation{Institut de Min\'eralogie, de Physique des Mat\'eriaux et
  de Cosmochimie (IMPMC), Sorbonne Universit\'e, CNRS UMR 7590, IRD
  UMR 206, MNHN; F-75252 Paris, France}

\date{\today}
\begin{abstract}
Optical conductivity measurements on a \bcs\ single crystal unveil an unusual linear behavior over a broad spectral range. In the paramagnetic phase above 300 K, the spectrum shows no gap, which contradicts the previously proposed scenario of a charge-transfer Mott insulator. \textit{Ab initio} dynamical mean field theory calculations including a retarded Hubbard interaction explain the data in terms of an incipient opening of a Co(3$d$)-S(3$p$) charge-transfer gap concomitant to incoherent charge transport driven by electronic correlations. These results point to a non-Fermi liquid scenario with Hund's metal properties in the paramagnetic state, which arises from an incipient Mott phase destabilized by low-energy charge fluctuations across the vanishing 3$d$-3$p$ charge-transfer gap. 
\end{abstract}
\maketitle

\section{Introduction}
\label{introduction}

The frequency dependent optical conductivity is a powerful probe of the deviations of charge transport properties of solids from the predictions of conventional band theory~\cite{basov2011}. The simplest form of this theory leads to the Drude model, which describes remarkably well the behavior of simple metals but also some strongly correlated systems such as heavy fermions with sizable mass renormalizations $m^*/m \sim 70$~\cite{Scheffler2005}. An anomalous metallic behavior is systematically observed in doped Mott or charge-transfer insulators, such as vanadates~\cite{vecchio2015,vecchio2017} and cuprate superconductors~\cite{uchida1991,takenaka2002}, characterized by a strong Coulomb repulsion between $d$ electrons. 
In these compounds, the Drude peak weight, the hallmark of coherent quasiparticle (QP) transport, reduces progressively as the Coulomb repulsion increases, with a resulting transfer of spectral weight from low to high energies. 
Recently, a novel phenomenology of anomalous charge transport has been reported in the so-called Dirac semimetals, where linearly dispersive electronic bands cross at the Fermi energy, $\varepsilon_F$, the so-called Dirac cones~\cite{Timusk_2013,Tabert_2016,carbotte2016}. In this case, the vanishing density of states at $\varepsilon_F$ suppresses the Drude peak and the linear band dispersion leads to a peculiar linear frequency-dependence of the real part of the optical conductivity, \resigw. Despite the absence of this peak, band theory and the concept of coherent QP remain valid.

\bcs\ is an antiferromagnetic (AF) insulator showing a N\' eel transition around 300~K \cite{mandrus_magnetism_1997,Fisher_1999}. It is an end member of the \bcns\ solid solution, where electronic correlations play a significant role \cite{hase_electronic_1995,klein2018importance}. Indeed, \citeauthor{takeda_transport_1995}~\cite{takeda_transport_1995} found a metal-insulator transition (MIT) in \bcns, concomitant to an antiferromagnetic to paramagnetic transition controlled by the Co/Ni substitution level, $x$. The resulting $T-x$ electronic phase diagram turns out to be similar to that of cuprates~\cite{keimer2015}, pnictides~\cite{paglione2010} and heavy fermions~\cite{gallagher2016}. A further similarity with cuprates and pnictides is a layered structure formed by conducting \ce{Co$_{1-x}$Ni_$x$S} square-lattice planes, separated by insulating BaS planes~\cite{grey_crystal_1970}. Although these similarities have stimulated several studies, the nature of the MIT remains controversial. Recently, the observation of a very large Rashba effect enhanced by the electric crystal field~\cite{santos-cottin_rashba_2016} renewed the interest in this system.

Very limited optical data exist
on this system. \citeauthor{kim1997}~\cite{kim1997} reported the signature of the MIT in \ce{BaCo_{0.9}Ni_{0.1}S_{1.9}} polycrystals. In that work, the lack of single-crystal data prevented the extraction of the intrinsic optical response and the study of the anisotropy. Later, \citeauthor{Takenaka_2001}~\cite{Takenaka_2001} investigated the effect of doping on the room temperature polarized optical conductivity of a series of \bcns\ single crystals at various doping levels up to $x = 0.28$. They proposed that \bcs\ is a Mott charge-transfer insulator with in-plane and out-of-plane gaps of 1 and 3.5 eV, respectively. Interestingly, their data indicate a large residual conductivity below these gaps, which raises questions about the unconventional properties of the insulating phase.

In the present work, we measure the optical properties of, and perform \textit{ab initio} dynamical mean field theory calculations for the strongly correlated material \bcs. We find that \resigw\ depends linearly on the frequency over a broad $\sim\ev{1}$ energy range. Our calculations qualitatively show that this linear \resigw\ and the absence of a Drude peak are not related to Dirac fermions but rather originate from a non-Fermi liquid state at the verge of a Mott metal-insulator transition. We find that the transport properties are poorly metallic, because of a vanishing charge-transfer gap between the cobalt $3d$ and the sulfur $3p$ bands, which contrasts with the previously proposed picture of a doping-driven MIT in \bcns~\cite{takeda_transport_1995}. Our results suggest a novel type of Fermi-liquid breakdown in a $d$-multiband system, where the tendency towards a Mott state is weakened by the low-energy charge fluctuations across the vanishing charge-transfer gap. We show that this situation is driven by retardation effects of the local Hubbard interaction, $U(\omega)$, which leads to a nearly half-filled configuration for a set of $d$ orbitals, and thus enhances the pivotal role of the Hund's coupling $J$ in the emergence of highly incoherent electronic properties. 

\section{Optical conductivity measurements}
\label{optical_conductivity_measurements}

We measured the in-plane near-normal-incidence reflectivity of freshly cleaved $ab$ surfaces of a high-quality \bcs\ single crystal grown by a self-flux method~\cite{Snyder_1994,Gelabert_1996}. The sample was mounted on the cold finger of an ARS Helitran cryostat. 
We employed a gold overfill technique~\cite{Homes_1993} to determine the absolute reflectivity with an accuracy better than 0.5 \%. The data were collected in the \mev{1.25}--\ev{1.5} range at eight temperatures from 5 to \K{320} using Bruker IFS113/v and 66/v interferometers. We complemented the data up to \ev{5.5} at room temperature using an AvaSpec 14x2048 optical fiber spectrometer. We determined the complex conductivity by employing a Kramers-Kronig transformation with a low-frequency Hagen-Rubens extrapolation. At high frequencies, we utilized the Tanner procedure~\cite{Tanner2015}, where the reflectivity is calculated from atomic X-ray scattering cross sections from 10 to \ev{60}, followed by a $1/\omega^4$ free electron termination. We merged the experimental and calculated reflectivity with a smooth cubic spline curve.

%

\begin{figure}[h]
	\includegraphics[width=\columnwidth]{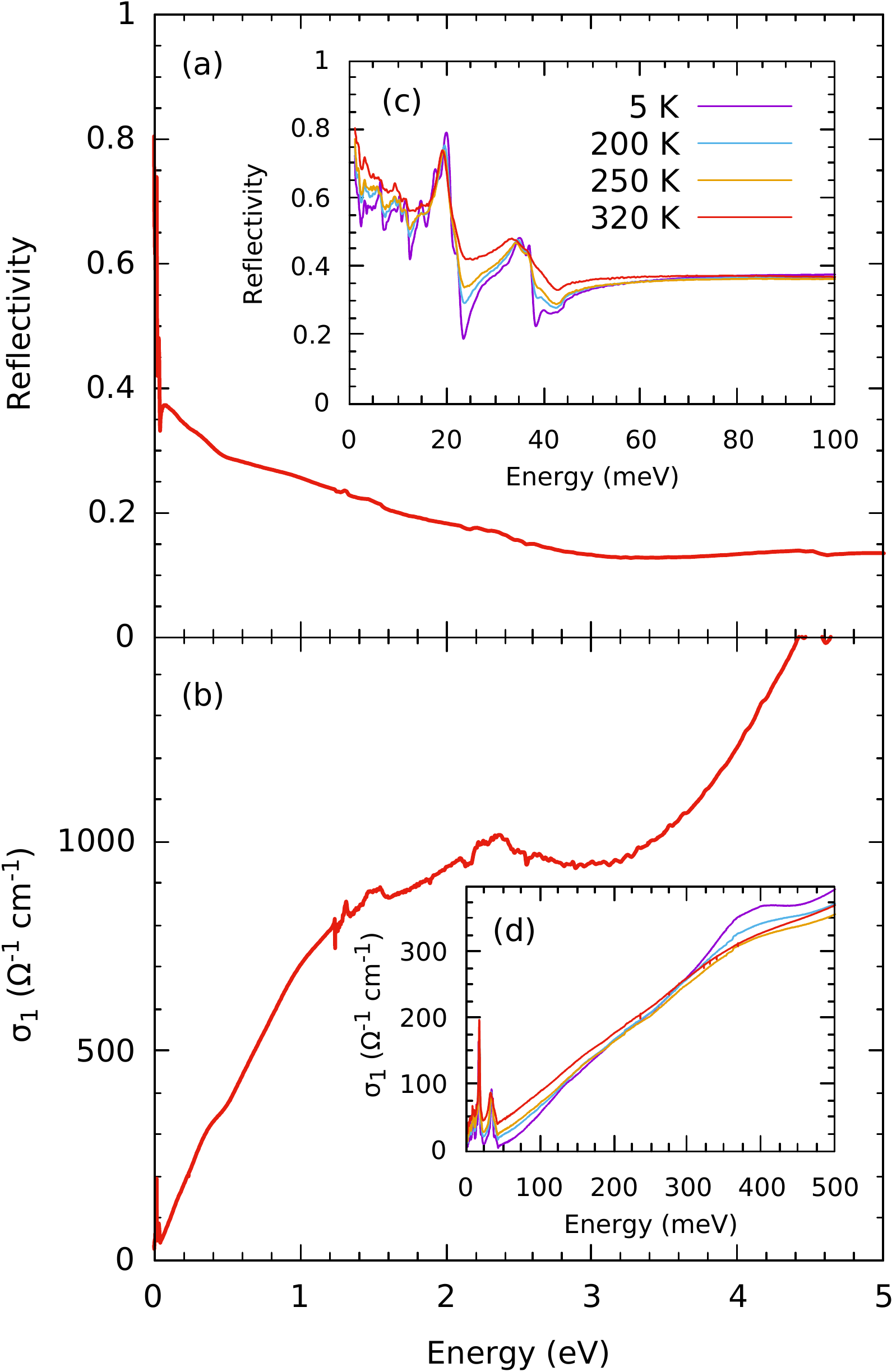}
  	\caption{(color online) (a) The reflectivity of \bcs\ in the paramagnetic state at 320~K and in (b) the corresponding real part of the optical conductivity. The inset (c) shows the temperature evolution of the reflectivity in the far-infrared. The sharp peaks are phonons, which get less screened by the electronic continuum at low temperatures. (d) The real part of the optical conductivity as a function of the temperature. In the antiferromagnetic state, a gap opens and low-energy spectral weight is transferred to about 400~meV. Note that at 320~K, the zero frequency extrapolation of \resigw\ is finite.}
  	\label{fig1}
\end{figure}

Figure \ref{fig1}(a) shows the in-plane reflectivity of \bcs~ measured at \K{320}. Figure~\ref{fig1}(b) shows the corresponding real part of the optical conductivity. Overall, our results agree with single crystal data reported by \citeauthor{Takenaka_2001}~\cite{Takenaka_2001}. Our crystals, however, show a higher dc conductivity, which is attributed either to a larger concentration of sulfur vacancies or to a higher purity. Note the absence of a conductivity peak at zero frequency and the strikingly linear dependence of the optical conductivity over a broad energy range $\sim$1 eV, which reflects a low mobility of the charge carriers. 

\begin{figure}[h]
  		\includegraphics[width=\columnwidth]{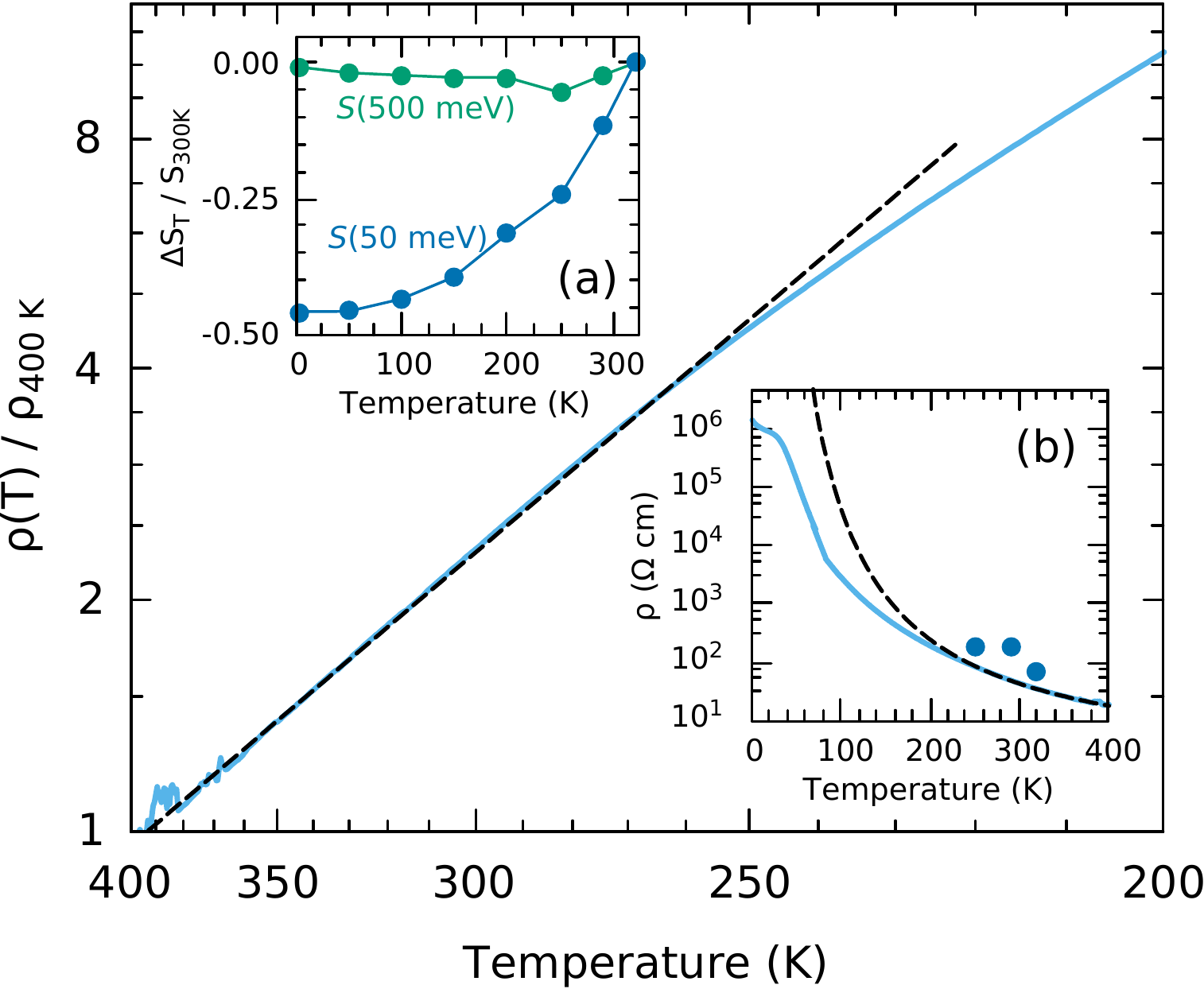}
\caption{(color online) The main panel shows a fit (dashed line) of the electrical resistivity (solid line) using a thermally activated model, as described in the text, which yields $E_a^\varrho = \mev{91}$. Note the logarithmic scale for $\varrho$ and the inverse temperature abscissas. Inset (a): cumulative spectral weight at \mev{50} and at \ev{0.5} [Eq.~(\ref{eq_sw})] with respect to the \K{320} value. Inset (b): dc electrical resistivity of \bcs\ compared to the dc resistivity obtained from the extrapolation to zero frequency of the optical conductivity (solid symbols). This extrapolation is reliable only at high temperatures, as explained in the text. The dashed line is the same fit shown in the main panel, stressing the deviation from a thermally activated regime below \K{200}.} 
  	\label{fig2}
\end{figure}

Fig.~\ref{fig1}(c) zooms into the low-energy reflectivity. The sharp peaks are polar phonons partially screened by the electronic continuum. This partial screening indicates the poor metallic character of the material. Upon decreasing the temperature, the phonons get sharper and better defined due to a decrease in the charge carrier concentration and hence a smaller screening. Panel (d) shows \resigw\ in the 0--\ev{0.5} range. At 320~K we have the aforementioned linear frequency dependence and a finite $\sigma_1(0)$ dc value. Below the N\'eel temperature a gap gradually appears below $\sim\mev{45}$ and it is almost fully open at 5~K. This gap opening is accompanied by a transfer of spectral weight from low energies to \ev{0.4}. Defining the cumulative spectral weight as 
\begin{equation}
S(\omega) = \int_0^\omega \sigma_1(\omega^\prime) d\omega^\prime \, ,
\label{eq_sw}
\end{equation}
we find that $S(\ev{0.5})$ is temperature independent within 5\%, as shown in Fig.~\ref{fig2}(a), indicating that the low-frequency spectral weight lost by the AF gap opening is fully recovered in the \ev{0.4} bump. 

We can relate the optical data to dc transport. Figure~\ref{fig2}(b) shows the dc resistivity of our sample compared to the zero energy extrapolation of \resigw. Our optical conductivity data allows for an accurate extrapolation only above 250 K, where both quantities are of the same order of magnitude. The temperature evolution of the reflectivity and the optical conductivity, as well as transport data, can be understood within a thermally activated behavior in the AF phase. This thermally activated charge transport scenario~\cite{Snyder_1994,Martinson_1996,Fisher_1999} becomes clear in the main panel of Fig.~\ref{fig2}, where the dc resistivity, $\varrho(T)$, is fitted to
\begin{equation}
\varrho(T) \propto \exp \left( \frac{E_a}{k_B T}\right) \, .
\label{eq_ta}
\end{equation}
A fit of the $\varrho(T)$ in the 250-400 K range yields an activation energy $E_a^\varrho = \mev{91}$. It is worth noticing, however, that the resistivity curve exhibits no true divergence at the lowest temperature measured of 1.8 K. Indeed, the activated exponential behavior stops at around 200 K, and it is no longer valid at lower temperatures, as more clearly depicted by the dashed line in Fig.\ref{fig2}(b). This inset also shows that the slope of the dc resistivity changes around 100 and, again, at 40 K, with a final saturation at the lowest temperatures.

Figure~\ref{fig2}(a) also shows the cumulative spectral weight up to \mev{50}, the region dominated by the AF gap. The AF gap progressively fills up with temperature, and vanishes in the paramagnetic phase, in agreement with neutron diffraction experiments\cite{mandrus_magnetism_1997}. 
This gap filling agrees with the thermally activated picture. However, the accuracy of the optical conductivity spectral weight is not enough to determine
the exact thermally activated regime temperature range, nor the activation energy.
Our activation energy is comparable but smaller than the value reported by \citeauthor{Fisher_1999}~\cite{Fisher_1999}, which may be ascribed to the presence of a small excess of sulfur, leading to a decrease of $E_a$. 

In any case, the important point here is to remark that the energy of both the AF gap ($\sim\mev{40}$) and the activation energy are much smaller than the proposed energy for a Mott gap in this material ($\sim\ev{1}$) \cite{Takenaka_2001}. Even more important, this AF gap is absent above room temperature, where the Mott insulating phase has been claimed. From now on, we are going to focus on the paramagnetic phase above 300 K, where our data does not show any optical gap.

The most peculiar feature of our data is the linear behavior of the room temperature conductivity over a broad energy range. \citeauthor{Takenaka_2001}~\cite{Takenaka_2001} attempted to provide a qualitative explanation in terms of a \ev{1} charge-transfer gap filled by a low DOS created by sulfur deficiency. However, the authors did not discuss this possibility in detail and made no conductivity calculations to explain the data. We stress the fact that the above linear behavior is highly anomalous and is observed in only a handful of compounds such as \ce{ZrTe5}~\cite{chen_optical_2015}, TaAs~\cite{xu_optical_2016,kimura_optical_2017}, the pyrochlore iridate \ce{E2Ir2O7}~\cite{sushkov_optical_2015}, and \ce{GaTa4Se8}~\cite{TaPhuoc2013}. 

\citeauthor{Timusk_2013} made analytical calculations within conventional band theory and argued that a linear dependence of the conductivity is the hallmark of Dirac cones in three dimensions~\cite{Timusk_2013,hosur_charge_2012,Tabert_2016,Ashby_2014,Neubauer_2016}. 
This scenario is clearly not applicable here, as Dirac cones at the Fermi level have been found neither in previous band structure calculations, nor in recent ARPES data~\cite{marino_unpublished}.

\section{\emph{Ab initio} DFT+DMFT calculations} 
\label{abinitio}

Previous calculations have been carried out within the local density approximation (LDA) supplemented by the Hubbard $U$ (LDA+U)~\cite{zainullina_ground_2011} or coupled with the dynamical mean field theory (LDA+DMFT)~\cite{zainullina2012}. These studies took $U =5$~eV as an empirical value for the static Hubbard repulsion. A comprehensive theoretical analysis is therefore required to unveil the non-trivial charge-transport mechanism in \bcs, which may also provide insights into the origin of the MIT in \bcns. 

With this aim, we carried out extensive \textit{ab initio} calculations of the optical conductivity of \bcs~by employing the LDA+DMFT+$U(\omega)$ approach. 
The technical details of these calculations are reported in the Supplemental Material\cite{supplemental_material}.
We utilized an atomic Slater parametrization of the Hubbard interaction, estimating $U$ and $J$ parameters from first principles with the constrained random phase approximation (cRPA)~ \cite{aryasetiawan2004,aryasetiawan2006}, in the so-called ``d-dp'' model~\cite{miyake2009,aichhorn2009,imada2010}. We computed the local Coulomb interaction involving the $d$-electrons, which is screened in a Wannier basis, by including both S($3p$) and Co($3d$) atomic orbitals in the low-energy window of the tight-binding model. Similar to the case of other transition metal compounds~\cite{werner2012,biermann2014,werner2016}, cRPA gives a strong frequency dependence of $U(\omega)$, which shows an abrupt reduction of the unscreened value at $\omega = 25$ eV, and yields a hopping renormalization factor $Z_B=0.55$~\cite{casula2012a,casula2012b}. 
The static limit, parametrized with spherical symmetry and averaged over the orbitals, corresponds to $U=\ev{2.92}$ and $J=\ev{0.90}$. 
We kept the full frequency dependence of the dynamic $U$ in our LDA+DMFT+$U(\omega)$ approach, by using a hybridization-expansion continuous-time quantum Monte Carlo impurity solver for density-density local interactions~\cite{werner2006,werner2010,gull2011}. 
To subtract correlation effects already included in the one-body LDA part of the Hamiltonian, we add to the self-energy a shift corresponding to the full-localized-limit double counting term $E_{DC}=-[U(n_\textrm{corr}-1/2)-J(n_\textrm{corr}/2-1/2)]$, where $n_\textrm{corr}$ is the self-consistent occupancy of the correlated $d$ manifold~\cite{aichhorn2011}. 

The main consequence of a dynamic $U$ is to significantly improve the screened values at low energies. Indeed, $U(0) \sim \ev{2.9}$ agrees with the experimental estimates provided in Ref.~\onlinecite{krishnakumar2001}, as opposed to the \ev{5} utilized in previous correlated calculations~\cite{zainullina_ground_2011,zainullina2012}. In order to stress the importance of a smaller static limit for $U$ while keeping 
the full frequency dependence we also ran standard LDA+DMFT calculations with 
a static $U$ matrix, without frequency dependence, utilizing the $\omega \rightarrow 0$ value of our dynamic calculation. 
All LDA+DMFT simulations are performed in the paramagnetic state, and the inverse electronic temperature has been set to $\beta= 1/k_BT$= 40 eV$^{-1}$.

%
\begin{figure}[t]
	\includegraphics[width=\columnwidth]{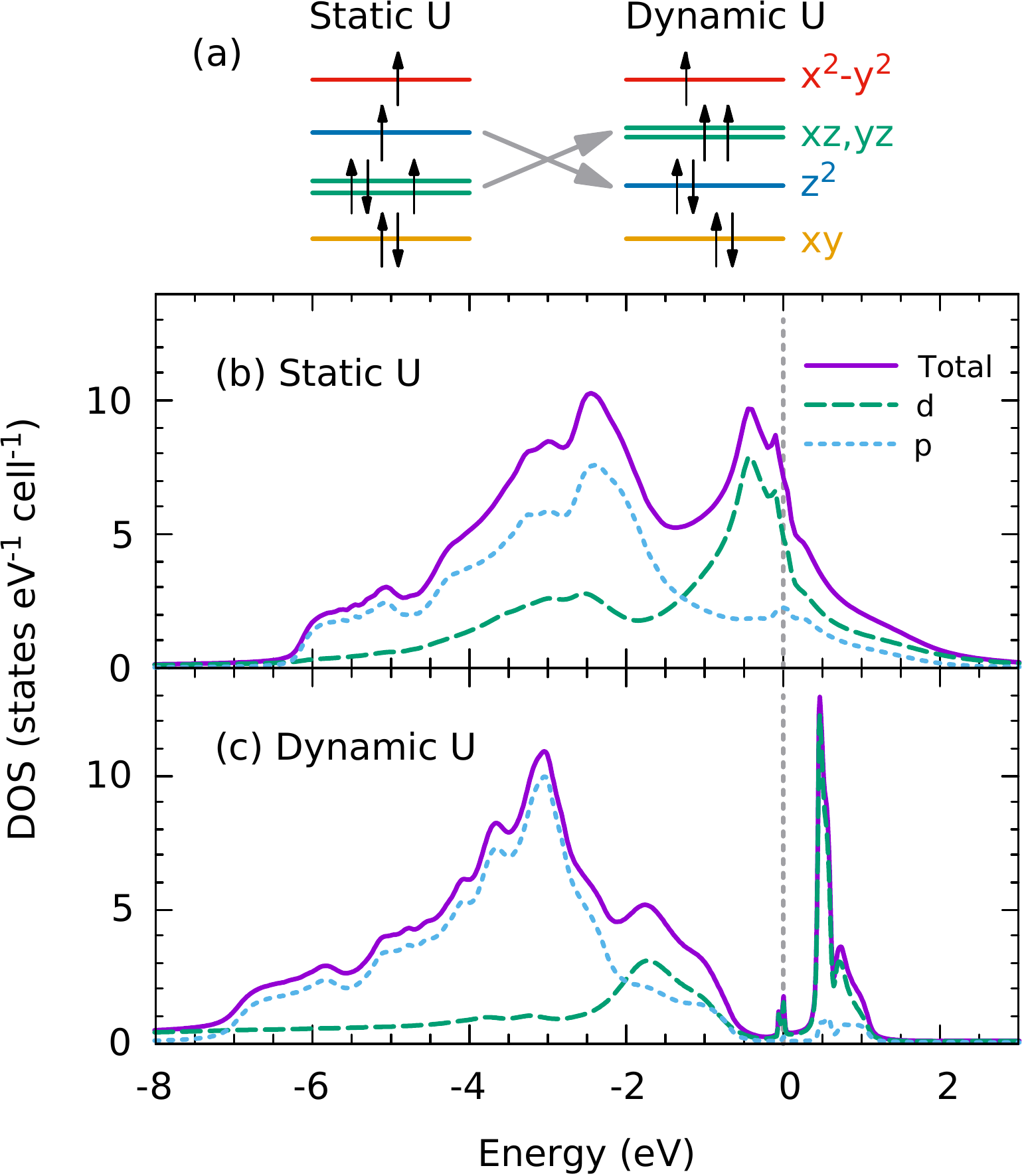}
  	\caption{(color online) (a) The hierarchy of the orbital occupations for a static or dynamic $U$, based on the crystal field theory for a square-pyramidal point group. Note that in the dynamic case $U(0)$ is the same as the value in the static $U$ calculation. 
  	(b) and (c) The $k$-integrated LDA+DMFT spectral functions (solid line) with static and dynamic $U$. The dashed lines are the partial densities of states (DOS) for $p$ (blue) and $d$ (green) orbitals.}
  	\label{fig3}
\end{figure}

Figures~\ref{fig3}(b) and (c) show the $k$-integrated spectral functions for both static and dynamic $U$. The difference between the two results is impressive. While the static $U$ solution is metallic with a large DOS at the Fermi level, the retarded interaction, while keeping the same static $U$ value at zero frequency, opens a small gap with the formation of in-gap states right at the Fermi energy. Even though we will discuss both approaches, we should note that only the dynamic $U$ results are compatible with the absence of a Drude peak in the optical conductivity of Fig.~\ref{fig1} and the small incoherent conductivity observed at low frequencies. Before showing more detailed calculations of \sigw, let us first look at the resulting orbital occupations, reported in Table~\ref{occupations}.

\begin{table}[h]
\caption{Occupations of $d$ orbitals per Co site from the LDA+DMFT
  solution with static and retarded Hubbard interactions. The total
  S($p$) and Co($d$) occupations are per unit cell.}
\label{occupations}
\begin{ruledtabular}
\begin{tabular}{ l d d }
\makebox[60pt][c]{Orbital} & \makebox[30pt][r]{Static $U$} &
\makebox[30pt][r]{Dynamic $U$} \\
\hline
\makebox[60pt][c]{$x^2-y^2$} & 1.17 & 1.12 \\
\makebox[60pt][c]{$z^2$}        & 1.36 & 1.94 \\
\makebox[60pt][c]{$xz$/$yz$} & 1.71 & 1.10 \\
\makebox[60pt][c]{$xy$}          & 1.93 & 1.99 \\
\hline
\makebox[60pt][c]{Total Co($d$)}    &15.76  & 14.50 \\
\makebox[60pt][c]{Total S($p$)}    & 22.24 & 23.50 \\
\end{tabular}
\end{ruledtabular}
\end{table}

The orbital occupations for static $U$ [Fig.~\ref{fig3}(a)] agree with crystal field theory for a square-pyramidal point group and with Hund's rules. The $xy$ orbital, aligned with the Co--Co direction, is the deepest and is fully occupied. Conversely, the $x^2-y^2$ orbital, which points to the Co--S--Co direction, is pseudogapped and nearly half-filled. The degenerate $xz$ and $yz$ orbitals are close to $3/2$ filling, while the occupancy for $z^2$ is a bit more than half-filled. Thus, the atomic energy level of the $z^2$ orbital is higher than the one of the $xz$/$yz$ orbitals. The levels are shuffled in the presence of stronger correlations induced by the dynamic $U$. The $z^2$ energy level sinks below $xz$/$yz$, and it gets almost fully-occupied, while the $xz$ and $yz$ become nearly half-filled. The competition between crystal field and local Coulomb repulsion brings about an atomic configuration that is prone to a Mott gap opening, with orbitals close to integer fillings. The resulting Mott gap is very small and it is filled by in-gap states. Indeed, the emergence of the Mott phase is not complete, with a residual hybridization of $p$ orbitals in the empty part of the spectrum. By increasing the strength of the correlations, the total $p$ and $d$ occupations approach their ionic model nominal value of 24 and 14, respectively (see Table~\ref{occupations}). This indicates that the system approaches, \emph{but does not reach}, a charge-transfer insulator state.

\begin{figure}[b]
  \includegraphics[width=\columnwidth]{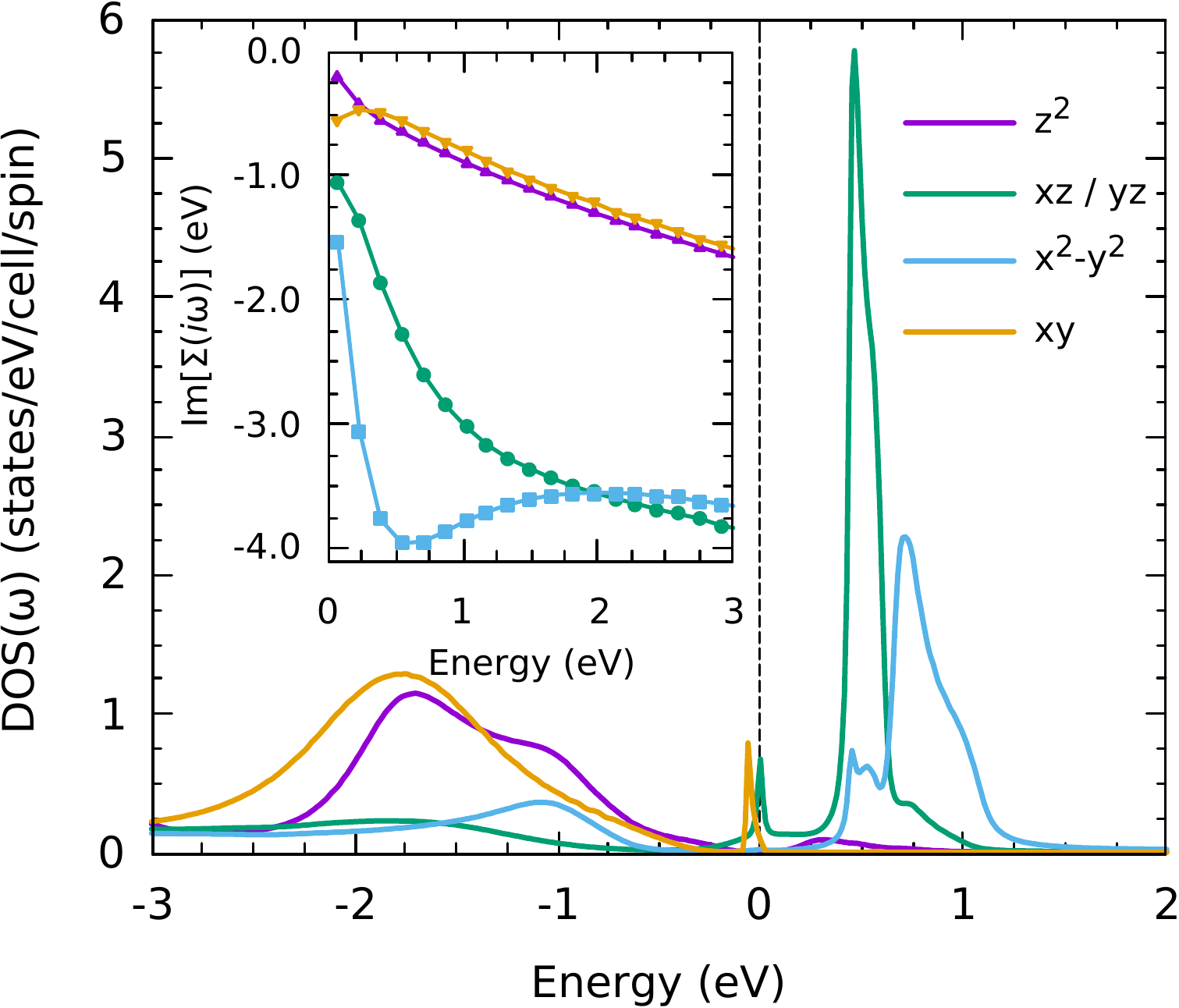}
  \caption{(color online) Main panel: orbital resolved DOS from LDA+DMFT+$U(\omega)$. Inset: imaginary part of the self-energy as a function of Matsubara frequencies obtained from LDA+DMFT+$U(\omega)$.
} 
 \label{Dos_and_Sw}
\end{figure}

To resolve the electronic properties of the residual charge carriers, we consider both the orbital projected DOS 
shown in the main panel of Fig.~\ref{Dos_and_Sw},
and the electron self-energies, whose imaginary part is plotted in 
the inset.
The in-gap states present at $\varepsilon_F$ have $xz$/$yz$ and $xy$ characters. The corresponding self-energies show strong correlation effects. 
At this temperature, they are 
not
compatible with a 
Fermi liquid.
Indeed, their linear extrapolation based on the smallest allowed Matsubara frequencies yields a finite value, while this must tend linearly to $0$ in a Fermi liquid. 
The behavior found here implies a strong incoherence, namely a very short lifetime, of the in-gap states, which must be reflected in experimental probes, such as the optical conductivity. 
Also note that the $x^2-y^2$ orbital shows enhanced correlations, whereas the $z^2$ is the only weakly correlated orbital. However, the latter two orbitals are not present at $\varepsilon_F$, thus they do not affect the transport properties at low energy, at variance with the $xz$, $yz$ and $xy$ orbitals.

\begin{figure}[h]
	\includegraphics[width=\columnwidth]{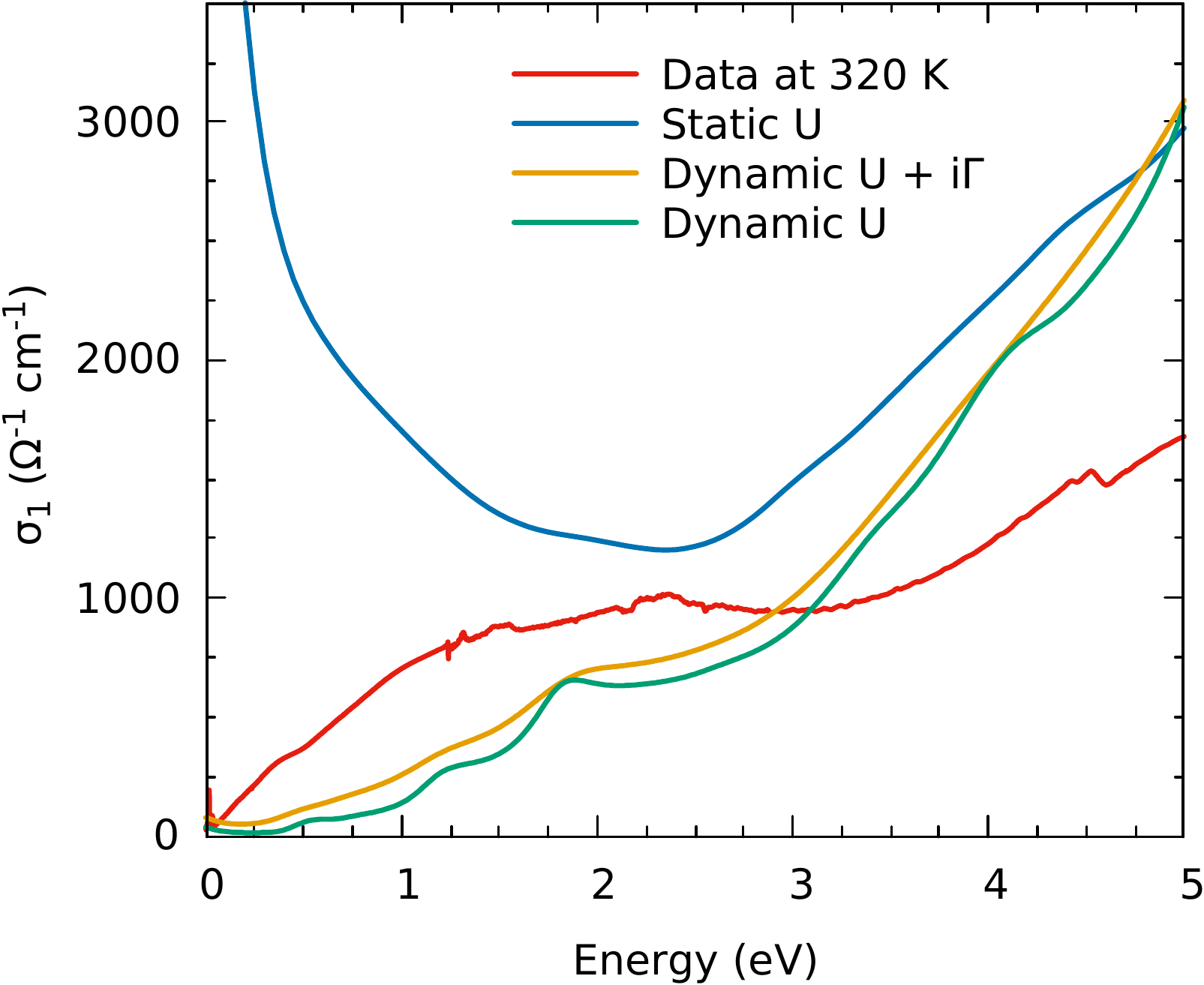}
	\caption{(color online) Experimental optical conductivity (red line)
		$\sigma(\omega)$ and the ones obtained from
		\emph{ab initio} calculations: both static and dynamic
		$U$ results are reported (blue and green, respectively),
		together with the case where we simulated the 
		scattering by impurities by adding a
		frequency-independent imaginary shift i$\Gamma$ to the self-energy,
		with $\Gamma=0.2$ eV (yellow curve). The tight-binding model has been
		extended up to 34 orbitals, by 
		including also the $s$ and $d$ Ba states. The self-energy used in
		the optical conductivity evaluation is the
		DMFT one at self-consistency, as obtained in the 22-orbital  $d-p$
		model. The calculation is performed on a  $60 \times 60 \times 60$
		$k$-points grid.}
	\label{optical_conductivity}
\end{figure}

We now study the implication of this peculiar electronic structure for the optical conductivity, computed in the Kubo linear-response formalism within the ``bubble'' approximation~\cite{pruschke1993,basov2011} by the generalized Peierls approach~\cite{tomczak2009}. 
The resulting in-plane optical conductivity is plotted in Fig.~\ref{optical_conductivity}, for both static and dynamic $U$. As expected, the correlated metal induced by the static Hubbard repulsion yields a Drude-like behavior, with a well defined Drude peak, totally absent in the experimental data.
In contrast, dynamic $U$ calculations give a behavior in semi-quantitative agreement with the experiment. 
A Drude-like response of the states at the Fermi level is suppressed by the incoherence associated with their non-Fermi liquid character, and with their small weight. 
We are left with an optical conductivity that vanishes for small energies with a linear behavior in the 0--1~eV range. However, the slope is smaller than in the actual experimental data. 

This discrepancy could be due to additional impurity scattering effects and/or to an even smaller Mott-Hubbard gap in the real material. 
To investigate the former hypothesis,
we added a frequency-independent imaginary shift to the DMFT self-energy. 
Upon increasing the shift amplitude, the low-energy slope of the optical conductivity increases, and the linear behavior becomes more pronounced (see Fig.~\ref{optical_conductivity}). 
Our \textit{ab initio} calculations find the main features of the optical conductivity and reproduce the linear \resigw, commonly assigned to three dimensional Dirac physics. However, the quantitative agreement remains poor, in particular the slope of the linear conductivity is underestimated with respect to the experimental outcome.

To explore the latter scenario, i. e. the presence of a smaller Mott-Hubbard gap in the real material, 
we derived a simplified DOS that allows one to manipulate the gap size and 
compute easily the optical conductivity in the approximation of momentum independent transition matrix elements.
An inspection of Fig.~\ref{fig3}(c) shows that, close to the Fermi level, the DOS $\rho(\omega)$ is composed of a broad continuum of filled stated; a small (incoherent) quasiparticle peak, and another peaked structure of empty states about \ev{0.5} above the QP peak. While the position of the incoherent QP peak is set by the charge-transfer gap, which is vanishing above 300~K, the position of the empty states, located at an energy $\Delta_2$ above the QP peak, is set by the underlying Mott-Hubbard gap. We found that our best description of the low-frequency experimental data through our simplified model produces $\Delta_2=\mev{75}$, compatible with the thermal activated behavior of the resistivity. The results of our model are reported in the Supplemental Material for both the room-temperature paramagnetic phase and the low-temperature AF gapped phase\cite{supplemental_material}. They point to an overestimation of the Mott-Hubbard gap in the LDA+DMFT+$U(\omega)$ calculations. This outcome is also in agreement with recent works\cite{van2014dynamical,PhysRevB.96.075102} that found a slight tendency of the dynamic $U$ to overestimate correlations in the LDA+DMFT framework.
We emphasize that, despite the uncertainty in the Mott-Hubbard gap magnitude, an underlying incoherent non-Fermi liquid is necessary to explain the optical conductivity anomalous features. This is provided by the proximity to a charge transfer regime and by a strong electron scattering, nicely captured by our \emph{ab initio} simulations, as shown in the inset of Fig.~\ref{Dos_and_Sw}.

%
%

\section{Conclusions}
\label{conclusions}

The optical conductivity of \bcs\ has  a remarkable linear dependence over a wide frequency range with no trace of a Drude peak. This peculiar behavior is particularly evident in the paramagnetic phase at high temperature, where the linear conductivity extrapolated to zero frequency remains finite, thus suggesting a picture of bad metal instead of that of a Mott insulator previously proposed. At low temperatures, in the zero-frequency limit, the low --- yet finite --- optical dc conductivity combined with a partial screening of the phonons indicates a thermally activated behavior with a small gap of $\sim \mev{40}$.

Our LDA+DMFT calculations from first principles with dynamic cRPA Hubbard interactions semiquantitatively reproduce the above anomalous features. We find that the dynamic $U$ plays a crucial role, for it brings the system into a strongly correlated regime, yielding a strong renormalization of the band structure, which accounts for the non-Fermi liquid behavior observed experimentally and, specifically, for the absence of the Drude peak. The method, however, slightly overestimates the magnitude of the underlying charge-transfer gap between Co($3d$) and S($3p$) states. A parametrization of the \textit{ab initio} 
DOS that allows this charge transfer gap to vary yields a quantitative agreement between calculations and experiment in both paramagnetic and antiferromagnetic phases at low energy.

The above experimental and theoretical results enable us to explain the charge dynamics of \bcs~ in terms of an incipient closing of this gap. Contrary to the previous case of Dirac semimetals, here the low-energy charge excitations across such a closing gap are incoherent because of the strong Hund's coupling inherent to the $3d^7$ shell of the Co$^{2+}$ ion. In the present case, dynamic correlations reshuffle the $d$ levels and drive them towards an effective three-orbital system, with a population close to 3 electrons, i.e. very close to half-filling, where the Hund's coupling effect is maximal. Because of the peculiar crystal field splitting, we argue that these excitations, favored by the net gain of kinetic energy associated with the larger $3p$ bandwidth, prevent the stabilization of the insulating Mott state. This conclusion suggests a novel type of Fermi-liquid breakdown in a $d$-multiband system driven by an interplay of charge-transfer gap and Hund's coupling.

%

\acknowledgments
PW acknowledges support from the Swiss National Science Foundation Grant 200021-140648 and 200021-165539.
~LdM is supported by the European Commission through the ERC-StG2016, StrongCoPhy4Energy, GA No724177. MC acknowledges the PRACE and GENCI allocations for computer resources under the project numbers PRA143322 and A0010906493, respectively. The DMFT calculations were run the the Brutus cluster at ETH Zurich and the Beo04 cluster at the University of Fribourg.
%
%
\bibliography{biblio}

\end{document}